\documentclass[12pt,preprint]{aastex}

\renewcommand{\vec}[1]{\mbox{\boldmath{$#1$}}}

\title{Solar cycle related changes at the base of the convection zone}
\author{Charles S. Baldner and Sarbani Basu}
\affil{Department of Astronomy, Yale University, P.O. Box 208101, New Haven, CT, 06520-8101}
\email{charles.baldner@yale.edu}

\begin{document}
\begin{abstract}
The frequencies of solar oscillations are known to change with solar 
activity.  We use Principal Component Analysis to examine these changes 
with high precision.  In addition to the well-documented changes in solar 
normal mode oscillations with activity as a function of frequency, which 
originate in the surface layers of the Sun, we find a small but 
statistically significant change in frequencies with an origin at and 
below the base of the convection zone.  We find that at 
$r=(0.712^{+0.0097}_{-0.0029})R_\odot$, the change in sound speed is 
$\delta c^2 / c^2 = (7.23 \pm 2.08) \times 10^{-5}$ between high and low activity.  
This change is very tightly correlated with solar activity.  In addition, 
we use the splitting coefficients to examine the latitudinal structure 
of these changes.  We find changes in sound speed correlated with 
surface activity for $r \gtrsim 0.9R_\odot$.
\end{abstract}

\section{INTRODUCTION}
\label{sec:intro}
Normal modes of oscillation of the Sun have provided a 
powerful tool to peer into the solar interior.  In particular, 
modern experiments, both ground- and space-based, have measured 
the intermediate degree global oscillation spectrum with high 
precision since the beginning of solar cycle 23.  Accurate 
determinations of interior structure and dynamics are now possible 
(see, e.g., review by \citealt{C-D02}).  These measurements 
contain a wealth of information about the fundamental causes of 
solar variability.  

It is generally believed that the seat of the solar dynamo is located 
at the base of the convection zone \citep[e.g., review by][]{Charbonneau2005}.  
Because helioseismology provides the only direct measurements of this 
region of the solar interior, these results can play an important role in 
constraining dynamo theories.  In particular, a number of authors have attempted 
to use global and local helioseismic techniques to determine limits on 
the strength of the magnetic field at the base of the convection zone 
\citep[e.g.,][and references therein]{Chouetal03}.  In this paper, we attempt to improve 
helioseismic measurements of changes in this region.

Global modes of solar oscillation are described by three numbers that
characterize the spherical harmonics that are used to define the horizontal
structure of the mode. These are (1) radial order $n$ that related to the number
of nodes in the radial direction, (2) the degree $\ell$ that is related to
the horizontal wavelength of the mode, and (3) the azimuthal order $m$
that defines the number of nodes along the equator.  In a spherically 
symmetric star, the $2\ell + 1$ modes of an $(n,\ell)$ multiplet are 
degenerate, but effects that break spherical symmetry such as magnetic fields or 
rotation lift the degeneracy and results in frequency splittings.  
The frequencies $\nu_{n\ell m}$ of the modes within a multiplet can be expressed
as an expansion in orthogonal polynomials:
\begin{equation}
\nu_{n\ell m}
= \nu_{n\ell} + \sum_{j=1}^{j_{\rm max}} a_j (n,\ell) \, {\cal P}_j^{(\ell)}(m).
\label{eq:acoefs}
\end{equation}
Early investigators \citep[e.g.,][]{Duvalletal86} commonly used Legendre
polynomials, whereas now one often uses the
Ritzwoller-Lavely formulation of the Clebsch-Gordan expansion \citep{RL91}
where the basis functions are polynomials related
to the Clebsch-Gordan coefficients.
In either case, the coefficients $a_j$ are referred to as $a$-coefficients 
or splitting coefficients.  Solar structure is determined by inverting 
the mean frequency $\nu_{n\ell}$, while the odd-order coefficients 
$a_1, a_3, \ldots$ depend principally on the rotation rate \citep{Durneyetal88}
and reflect the advective, latitudinally symmetric part
of the perturbations caused by rotation.  Hence, these
 are used to determine the rate of rotation inside the Sun.
The even order $a$ coefficients on the
other hand result from
magnetic fields and asphericities in solar structure,
and the second order effects of rotation
\citep[e.g.,][]{GT90,DG91}.

Solar oscillation frequencies are known to vary on timescales 
related to the solar activity cycle.  This was first suggested 
by \citet{WN85} and confirmed soon after by \citet{Elsworthetal90} 
and \citet{LW90}.  It was quickly established 
that the frequency shifts were strongly correlated with surface activity 
\citep[etc.]{Woodardetal91,BB93,Elsworthetal94,Reguloetal94}.  \citet{LW90} 
observed that the frequency shifts depended very strongly on mode 
frequency $\nu$, and very weakly on degree $\ell$ of the mode, and 
\citet{A-Getal92} and \citet{Elsworthetal94} confirmed these results.  
These authors concluded that all or most of the physical changes 
responsible for the changes in frequency were confined to the 
shallow layers of the Sun.  In general, this picture has been 
confirmed in more recent studies (e.g., observational 
results:  \citealt{HKH99,HKH02,BA00,Verneretal04,DG05}, etc., and theoretical 
results:  \citealt{Goldetal91,Balmforthetal96,Lietal03}, etc.).  A change in
the second helium ionization zone at $r=0.98R_\odot$, first suggested by \citet{Goldetal91} 
and \citet{Gough02}, has been confirmed by \citet{BM04} and 
\citet{Verneretal06}.  

The even-order mode splitting parameters sample effects of 
structural asphericities on the mode 
frequency.  \citet{K88} suggested that they were correlated with 
observed changes in surface temperature.  Subsequent work has 
shown that the aspherical components of the mode frequencies 
are tightly correlated with surface magnetic activity 
\citep{HKH99,Antiaetal01}.  This high correlation 
lends further credence to the idea that frequency shits are 
caused by surface and/or near-surface effects.  This can be tested 
directly with high degree modes that sample the near-surface 
layers of the Sun.  However, as the degree $l$ increases, global 
modes become increasingly hard to measure precisely due to the 
decrease in mode lifetimes \citep{Rhodesetal98,R-Setal01,Korzenniketal04}.  
The lack of reliable measurements of these modes has led some authors 
to use ring diagrams to measure high degree modes and measure changes 
in the shallow layers of the solar convection zone.  These studies have 
confirmed that structural changes do occur in the near-surface layers of the 
Sun \citep{Betal07}.

Direct inversions of changes 
in the structure of the solar interior probed by the spherically symmetric global 
modes have not yielded any measurable differences in the deep 
interior \citep{Basu02,E-Detal02} and there have been 
upper limits placed on the changes at the base of the convection 
zone \citep{E-Detal02}.  \citet{CS05} and \citet{SC05} have presented 
evidence of a possible change in mode frequencies with lower turning 
points near the base of the convection zone.

Internal dynamics, on the other hand, show clear and unequivocal 
evidence of change over the course of the solar cycle.  
In the convection zone, bands of different rotational velocities 
(called zonal flows) have been shown to migrate poleward at 
high latitudes and equator-ward at low latitudes 
\citep{Schou99,Howeetal00,AB00,AB01}.  Temporal variations in dynamics 
have been shown to penetrate throughout the entire convection 
zone, and even below \citep{Vorontsovetal02,BA06,Howeetal05,Howeetal06}.

One conclusion that can be drawn unequivocally from previous 
studies of the changes in solar structure is that any changes 
deeper than those in the outermost layers of the Sun are very small, and hence 
very difficult to detect through their signatures in oscillation 
frequencies.  \citet{CS05} used a smoothing technique 
to attempt to remove the effect of surface variations, and 
found that the scaled frequency differences showed evidence 
of change near the base of the convection zone, but could not 
say more about the physical nature of the changes.  Attempts 
to invert the frequencies directly have never shown any changes 
larger than the inversion errors \citep[e.g.,][]{Basu02,E-Detal02}.  
Therefore, although current 
helioseismic instruments have determined the solar oscillation 
frequencies with tremendous precision, statistical errors in 
those frequencies are still too large to make any direct detections 
of structure changes in the deep interior.

In this work, we take a somewhat different approach to attempting to 
detect changes at the base of the convection.  
We use Principal Component Analysis (PCA) to separate the frequency 
differences over the last solar cycle into a linear combination of 
different time-dependent components.  This has the effect of 
decreasing the effects of 
measurement errors in the measured helioseismic frequencies, which 
allows us to isolate as precisely as possible the changes in frequency 
over time.  In section \ref{sec:data}, 
we describe the data used, and the methods employed to analyze 
them.  In section \ref{sec:results}, we present the results in detail.  
Finally, in section \ref{sec:conc} we discuss the significance of 
the results and presents our conclusions.

%
\section{DATA \& ANALYSIS}
\label{sec:data}
\subsection{Data}
For this work, we use helioseismic global-mode data sets from two 
different projects, one from the 
Michelson Doppler Imager (MDI) on-board the SOHO spacecraft, and the other 
from the Global Oscillations Network Group (GONG).  The MDI mode sets 
consist of frequencies and splittings obtained from 72-day long time 
series \citep{Schou99}.  We use 54 of these sets, spanning the period from  1996 May 1 
to 2007 May 16.  The GONG mode sets are derived from 108-day time series \citep{Hilletal96}.  
Although GONG provides sets that overlap in time, we only use non-overlapping 
sets in the present work.  We use 40 sets from the period 1995 May 7 to
2007 April 14.  Because the two sets are from completely different 
instruments and independent data reduction pipelines, any real solar 
signatures should show up in both sets.  The $f$-modes do not sample the 
deep interior and are dominated by surface effects, so we exclude them 
from our study.  The $n = 1$ modes have larger errors than the higher 
order modes, and so we exclude them as well.  The included modes are 
low and intermediate degree modes up to $\ell=176$, with order $n$ from 
2 to 16.

As a proxy for total solar activity, we use the 10.7cm radio flux 
measurements taken by the Dominion Radio Astrophysical Observatory (DRAO)
\footnote{Data can be found at \\
http://www.drao-ofr.hia-iha.nrc-cnrc.gc.ca/icarus/www/sol\_home.shtml}.  
This measurement has been found to be very tightly correlated with 
solar activity \citep[e.g.,][]{Tapping87}.  We average the $F_\mathrm{10.7}$ 
measurements over 72-day periods 
for comparison with MDI data, and 108-day periods for comparison with GONG 
data.  For latitudinal structure in surface activity, we use the surface 
magnetic field, taken from SOHO/MDI synoptic maps
\footnote{MDI synoptic maps of Carrington rotations can be found at\\
http://soi.stanford.edu/magnetic/index6.html}.  
The magnetic field strengths are averaged over the same 72-day periods 
as the $F_{10.7}$ data, and over the appropriate ranges in latitude.

\subsection{Method}
We use Principal Component Analysis to describe the temporal variations of the 
weighted frequencies as a small number of uncorrelated basis functions.  The use of 
PCA is a common technique in multivariate data analysis to reduce dimensionality 
and expose underlying variables (see \citealp[Chapter 2]{MH87}, 
for a discussion of its astrophysical applications).  
A brief description of the method and discussion of its limitations is 
included in the Appendix.  PCA is a technique whereby a set of observations 
is expressed as a set of uncorrelated vectors.  
The usefulness of the technique arises from 
the fact that variation of the data about the first vector is maximal, 
and about the second vector, maximal subject to the constraint that it 
be orthogonal to the first vector, and so on.  In other words, PCA 
provides a very efficient linear representation of a data set, and it 
is able to substantially reduce the dimensionality of the data set 
without losing any significant information.  

It has been known for many years that the frequencies of the solar global 
modes of oscillation change with the solar activity cycle.  With the 
arrival of high quality measurements of intermediate degree modes, it 
is clear that the amount of frequency shift over the solar cycle is 
dependent on the mode.  Each mode has an associated mode inertia $E_{nl}$.  
Frequency differences can be scaled by the quantity $Q_{n\ell}=E_{n\ell}/\bar{E}_0(\nu_{n\ell})$, 
where $\bar{E}_0(\nu_{n\ell})$ is the inertia of the $\ell = 0$ modes 
interpolated to the frequency of the $(n,\ell)$ mode 
\citep{C-DB91}.  This scaling accounts for 
the fact that the frequencies of modes with lower inertia are changed by a 
larger amount than modes with a higher inertia by the same underlying perturbation.  
When scaled in this way, the degree-dependence of the frequencies over the 
solar cycle largely vanish, and the frequency changes become 
slowly varying functions of frequency only \citep[e.g.,][]{Chaplinetal01,Basu02}.  

Our data points are the scaled frequency differences 
$Q_{n\ell}\delta \nu_{n\ell} / \nu_{n\ell}$.  For the MDI observations, there 
are 54 total mode sets in our work, which, when one is removed to be used 
as the base set, giving us 53 sets of scaled frequency differences.  There are 40 
mode sets in our GONG data set, or 39 difference sets.  Because the PCA method 
requires a complete covariance matrix (see the Appendix), we 
can only include 
a mode if it is present in all mode sets.  This dramatically reduces the 
amount of usable information, particularly since many excluded modes are 
missing in only one or two sets out of 53.  For these modes, it is possible to 
interpolate a value for the missing frequency differences.  
In this case, because most of the frequency differences for the mode 
in question will be actually observed, any errors 
introduced through the interpolation will have a negligible effect on the 
PCA results.  We tested two interpolation methods --- one a spline interpolation 
along ridges in the $\ell$-$\nu$ diagram (interpolated from modes with the 
same radial order $n$), and the 
other a linear interpolation between neighboring modes in time.  When tested 
against existing modes, the interpolation along time proved superior, reproducing 
the actual data to better than a factor of 1.2$\sigma$, 
while the interpolation along the ridge results had a 2$\sigma$ distribution.  
Therefore, only results using the time interpolation are discussed in this paper, but the 
PCA results using interpolation over degree were entirely consistent.  Finally, to ensure 
that the PCA is robust, Monte Carlo simulations were performed to ensure that 
the exclusion of certain modes would not affect the results.  The PCA analysis 
of these data appears to be very robust.  Errors in the PCA components were 
computed by means of a Monte Carlo simulation.

In addition to the mean frequencies $\nu_{n\ell}$, which contain information about 
the spherically symmetric part of the solar interior, we have the even-order splitting 
coefficients $a_{2j}(n,\ell)$, which allow us to reproduce mode frequencies as a 
function of latitude.  Latitudinal frequencies as a function of colatitude 
$\theta$ can be obtained as follows:
\begin{equation}\label{eq:lat}
\nu_{n\ell}(\theta) = \nu_{n\ell} + \sum_k \frac{\ell a_{2k}(n,\ell)}{\mathcal{Q}_{\ell k}}P_{2k}(\cos \theta),
\end{equation}
where $\mathcal{Q}_{\ell k}$ are the angular integrals given by
\begin{equation}
\int_0^{2\pi} d\phi \int_0^\pi \sin \theta d\theta Y_\ell^m(Y_\ell^m)^* P_{2k}(\cos \theta) = \frac{1}{\ell}\mathcal{Q}_{\ell k}\mathcal{P}_{2k}^{(\ell)}(m),
\end{equation}
and $P_{2k}(\cos \theta)$ are Legendre polynomials of degree $2k$, and the 
$\mathcal{P}_{2k}^{(\ell)}$ are the same polynomials as in equation \ref{eq:acoefs}.

Ultimately, we are interested in helioseismic data for what they can 
tell us about the solar interior.  To extract this information, we invert 
these data sets for the parameters of interest.  We 
use two different inversion techniques, and invert for the change in 
sound speed relative to the comparison frequency.  The first technique used 
is Subtractively Optimized Local Averages \citep[SOLA][]{PT94}.  
A description of the implementation used here, and how to select 
inversion parameters can be found in \citet{R-Setal99}.  The 
second inversion technique is that of Regularized Least Squares (RLS).  The 
implementation used here and the selection of inversion parameters have been 
discussed in \citet{AB94} and \citet{BT96}.  The use 
of these two techniques in tandem is crucial because the techniques are 
complimentary in nature \citep{Sekii97}.  Inversions can be trusted if 
both inversion techniques return the same results.

The quantities which we invert are appropriately scaled eigenvectors from 
the PCA of the frequency differences.  The differences are taken relative to 
some fiducial set, usually the first mode set (corresponding to activity minimum 
at the beginning of solar cycle 23).  The eigenvectors, $\vec{\xi}_i$, are 
normalized so that $\vec{\xi}_i \cdot \vec{\xi}_i = 1$.  Each data set is 
the vector $\vec{x}(t)$ that contains the individual mode frequencies 
$Q_{n \ell}\delta \nu_{n \ell} / \nu_{n \ell}$.  These data can be 
represented as a linear combination of the the eigenvectors with
coefficients given by $c_i(t) = \sum_j \xi_{ij}x_j(t)$ (see the Appendix).  
The scaled eigenvector $c_i(t)\vec{\xi}_i$, therefore, has a physically reasonable 
magnitude, and in all following cases, it is this quantity that we invert.  
In general, we choose the coefficient with the largest magnitude, which represents 
the largest variation in sound speed.  This coefficient is usually the one associated 
with the set at maximum activity.

\section{RESULTS \& DISCUSSION}
\label{sec:results}
\subsection{Mean frequencies}
\begin{figure}
\plotone{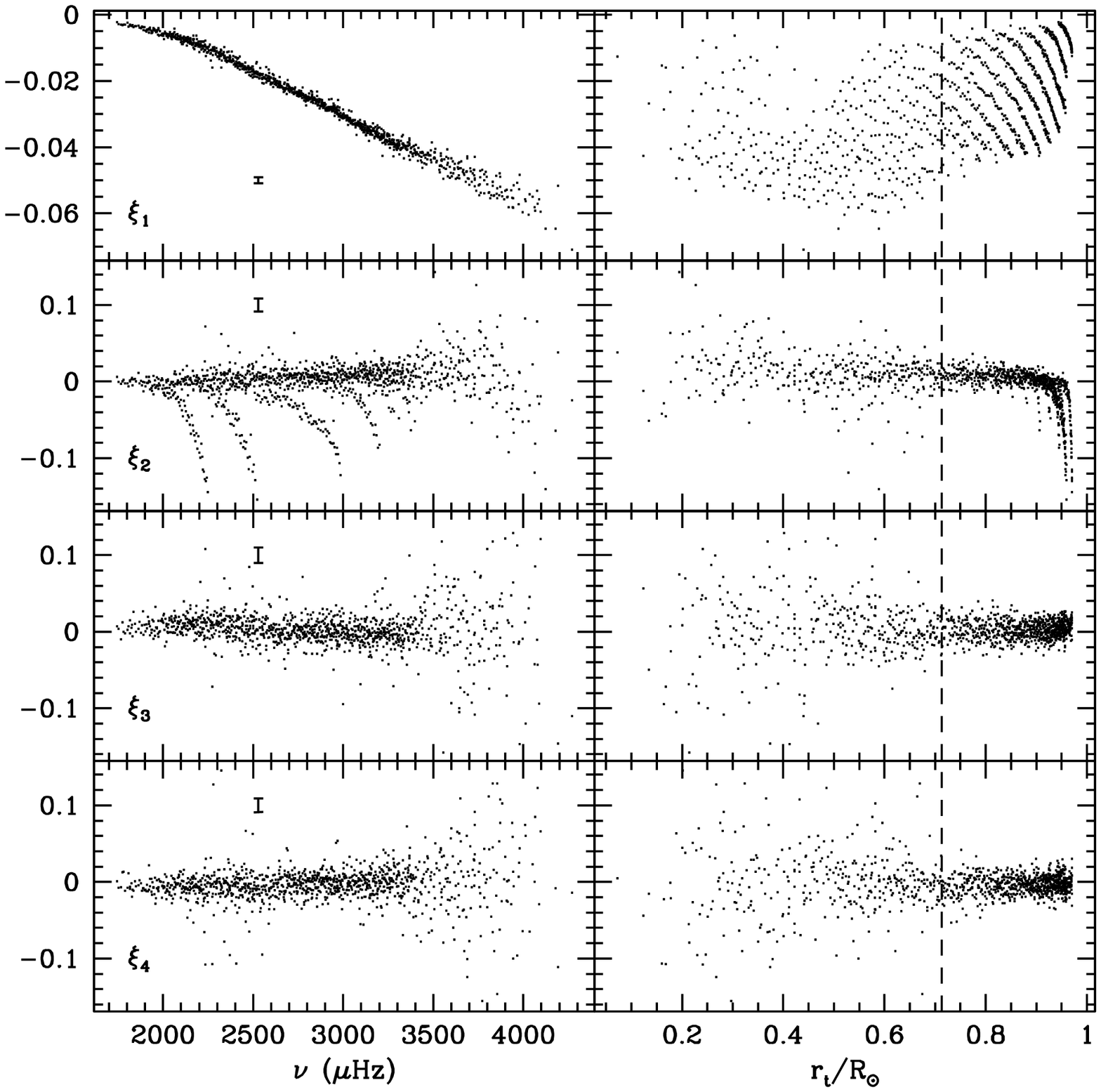}
\caption{The first four eigenvectors for the MDI data set are shown.  The 
base set is \#1216 (1996 May 1).  The left-hand panels show the eigenvectors 
as a function of frequency, and the right-hand panels show it as a function 
of the lower turning point of the mode.  The vertical line shows the 
position of the base of the convection zone.  The vertical axis units are 
arbitrary (the vectors are normalized so that $\vec{\xi}_i \cdot \vec{\xi}_i=1$).  
The error bars show representative errors calculated using a Monte Carlo 
simulation.
\label{fig:MDIvec}}
\end{figure}

\begin{figure}
\plotone{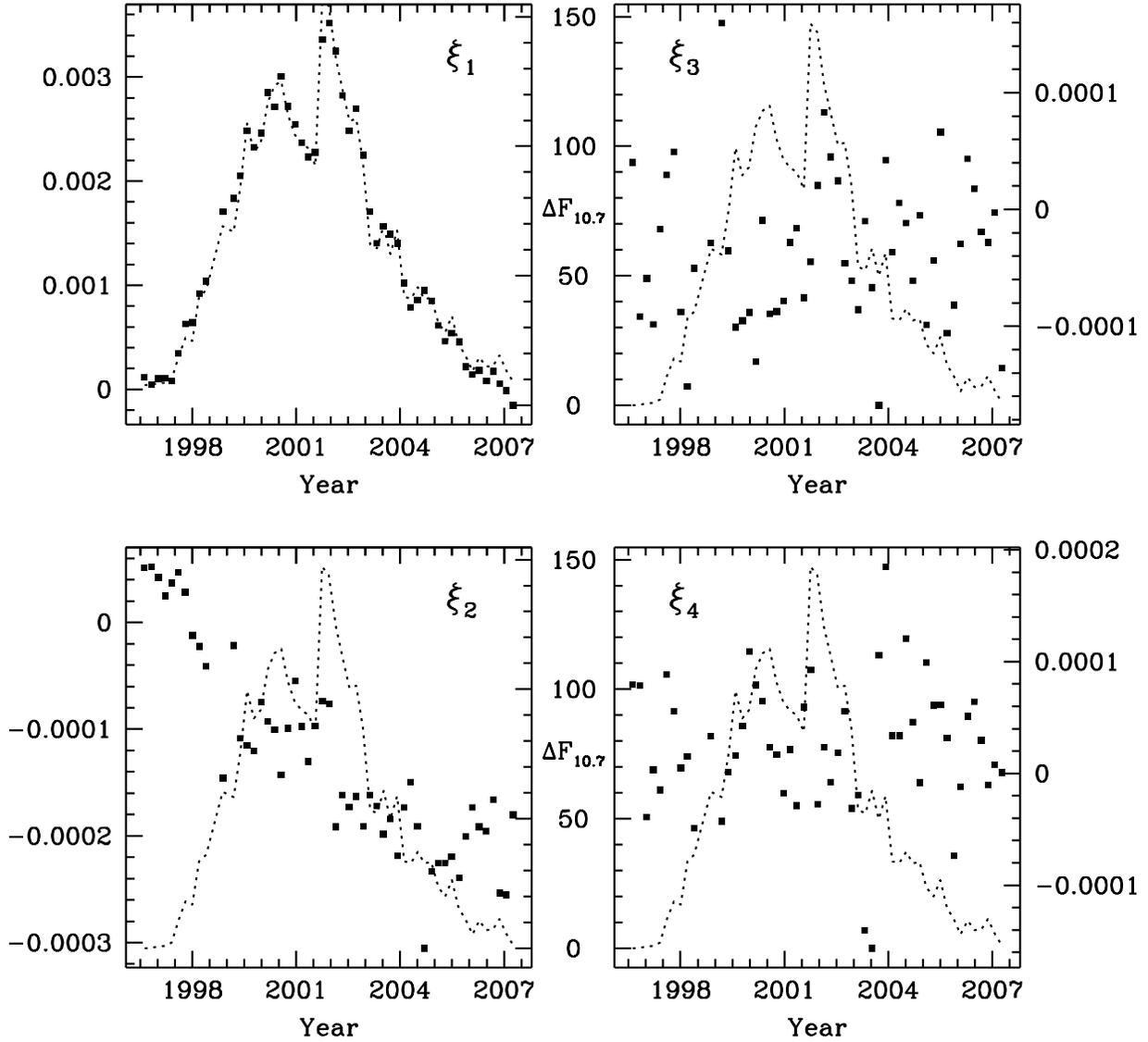}
\caption{The scaling coefficients for the first four eigenvectors are shown as 
points.  They are shown as a function of time (the start date of the 
MDI mode sets).  The dotted line is the change in 10.7cm radio flux with 
respect to the beginning of the solar cycle.  The units are Solar Flux 
Units (SFU).
\label{fig:MDIcoef}}
\end{figure}

We have performed the PCA decomposition of the MDI mean frequencies with respect 
to the first set (set \#1216, start day 1996 May 1, end day 1996 July 12).  This 
is a low activity set.  The first four eigenvectors $\vec{\xi}_1$ --- $\vec{\xi}_4$ 
are shown in Fig. \ref{fig:MDIvec}, both as a function of frequency and of the 
lower turning point of the modes.  
The scaling coefficients for the first four eigenvectors are presented in Fig. 
\ref{fig:MDIcoef}.  Also shown in the figure is the difference in the 10.7 cm 
radio flux between the first set and the subsequent sets.

When plotted as a function of frequency, the first eigenvector $\vec{\xi}_1$ 
appears to be almost entirely due to near-surface effects --- it is seems to 
be a slowly varying function of frequency 
only.  As a function  of $r_t$, however, a change in the average level 
of the frequency differences can be seen below the base of the convection zone 
(around $r \approx 0.713R_\odot$ --- marked by a vertical line in 
Fig. \ref{fig:MDIvec}).  This implies a time 
dependent change near the base of the solar convection zone.  We conclude, therefore, 
that there is a statistically significant component of the frequency 
variability picked out by $\vec{\xi}_1$ that does not originate at the surface.
The coefficients for $\vec{\xi}_1$ are tightly correlated with the 10.7cm 
radio flux, a proxy for surface activity; the correlation coefficient is 0.99.  
A linear regression fit to the change in radio flux relative to solar minimum, 
$\Delta F_\mathrm{10.7}$, gives the relation between the 10.7cm flux and the 
coefficients of $\vec{\xi}_1$ as:
\begin{equation}
\label{eq:f10line}
c_1 = (2.53\times 10^{-5} \pm 3.13 \times 10^{-6})\Delta F_\mathrm{10.7} + 3.95\times 10^{-5} \pm 1.6 \times 10^{-4}.
\end{equation}
Thus, all changes, as manifest by $\vec{\xi}_1$, are tightly correlated with 
surface magnetic activity.

The second eigenvector, $\vec{\xi}_2$, also shows variability over the solar cycle.  
Comparing it as a function of $\nu$ and of $r_t$, it is clear that the structure is neither 
a pure function of frequency nor of lower turning point.  The first four obvious 
downturn features correspond to modes of order $n=2$, 3, 4, and 5.  As a 
function of $r_t$, it seems clear that the structure is concentrated at or 
near the surface.  However, the differences can not be fit by the usual ``surface 
term", and further examination reveals that it cannot be fit even by higher-order 
surface terms of the form considered in \citet{BV93} and \citet{A95}.  The coefficients for 
$\vec{\xi}_2$ offer some hint as to what is going on.  They exhibit no obvious 
solar cycle dependency, but rather seem to be a roughly linearly decreasing 
function of time.  There does not seem to be any periodicity on a scale of 
eleven years or shorter.  Larson \& Schou (2008) have undertaken an in depth study of the
systematics in the MDI data reduction pipeline.  The plate scale in the
MDI instrument has changed slightly over the course of SOHO's mission,
and they have shown that the effect of the resultant
error in the measured radius introduces errors that look exactly like
the $\vec{\xi}_2$ eigenvector computed in this work.  We do not believe, therefore, 
that the features in $\vec{\xi}_2$ are solar in origin, but are rather artifacts 
from the MDI data reduction pipeline.  As we will show below, analysis of 
GONG data confirms our belief.

\begin{figure}
\plotone{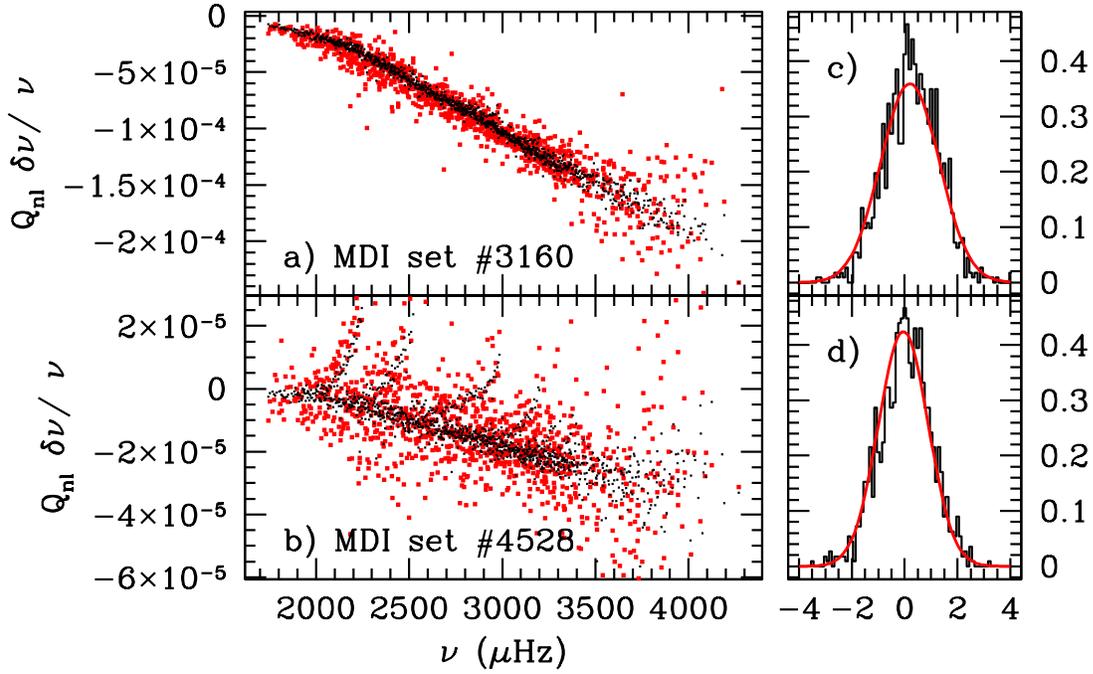}
\caption{To demonstrate that the PCA representation of the data does indeed 
capture the actual data, we present two reconstructed data sets and compare 
them to the actual data.  Two frequency difference sets are examined:  sets \#3160 
(2001 August 27, panel a) and \# 4528 (2005 May 26, panel b), both with respect 
to \# 1216.  At left, the frequency differences are plotted (panel (c) for set 
\#3160 and panel (d) for set \#4528).  The actual data is shown 
in red, and the reconstructed data set using $\vec{\xi}_1$ and $\vec{\xi}_2$ 
are overplotted in black.  The normalized residuals are binned and shown 
against a Gaussian curve to show that the information which has been 
removed is statistically random.
\label{fig:MDIrecon}}
\end{figure}

The third and fourth eigenvectors, $\vec{\xi}_3$ and $\vec{\xi}_4$ do not exhibit 
any significant structure at all.  The vector $\vec{\xi}_3$ shows a slight trend 
with frequency, but the scaling coefficients $c_3$ are normally distributed, and 
we cannot identify any physical significance in this eigenvector.  The remaining 
eigenvectors are statistically consistent with Gaussian noise distributed around zero.
We conclude, therefore, that the temporal variation of the MDI frequencies 
is dependent on a linear combination of $\vec{\xi}_1$ and $\vec{\xi}_2$ alone.  
In Fig. \ref{fig:MDIrecon}, we show two data sets reconstructed from the 
first two eigenvectors.  This figure 
shows that the PCA decomposition does indeed accurately capture the original data 
while significantly reducing the random scatter in the data.  The residuals 
normalized by the errors are plotted, and are consistent with Gaussian noise, with 
distributions of 1.1$\sigma$ and 0.9$\sigma$ for the two cases.  Having confirmed that 
the third and subsequent eigenvectors are Gaussian noise, we do not consider them 
further in this paper.  
This reduction in noise is important for attempting to invert the small signatures 
we are looking at here.

The fact that the PCA is applied to a set of mode sets relative to a single 
base set raises the possibility that we are unduly influenced by the choice 
of that base set.  We therefore repeat the PCA taking a base set from 
halfway up the solar cycle:  MDI set \#2224 (start date: 1999 February 3, end date 
1999 April 16, and an activity level during the 72 day period of $F_\mathrm{10.7}=130.7$ 
SFU).  The eigenvectors are consistent with those obtained from the base 
\#1216 set insofar as their inner products are close to unity. We conclude, therefore, 
that the PCA results are not unduly influenced by the choice of base set.

\begin{figure}
\plotone{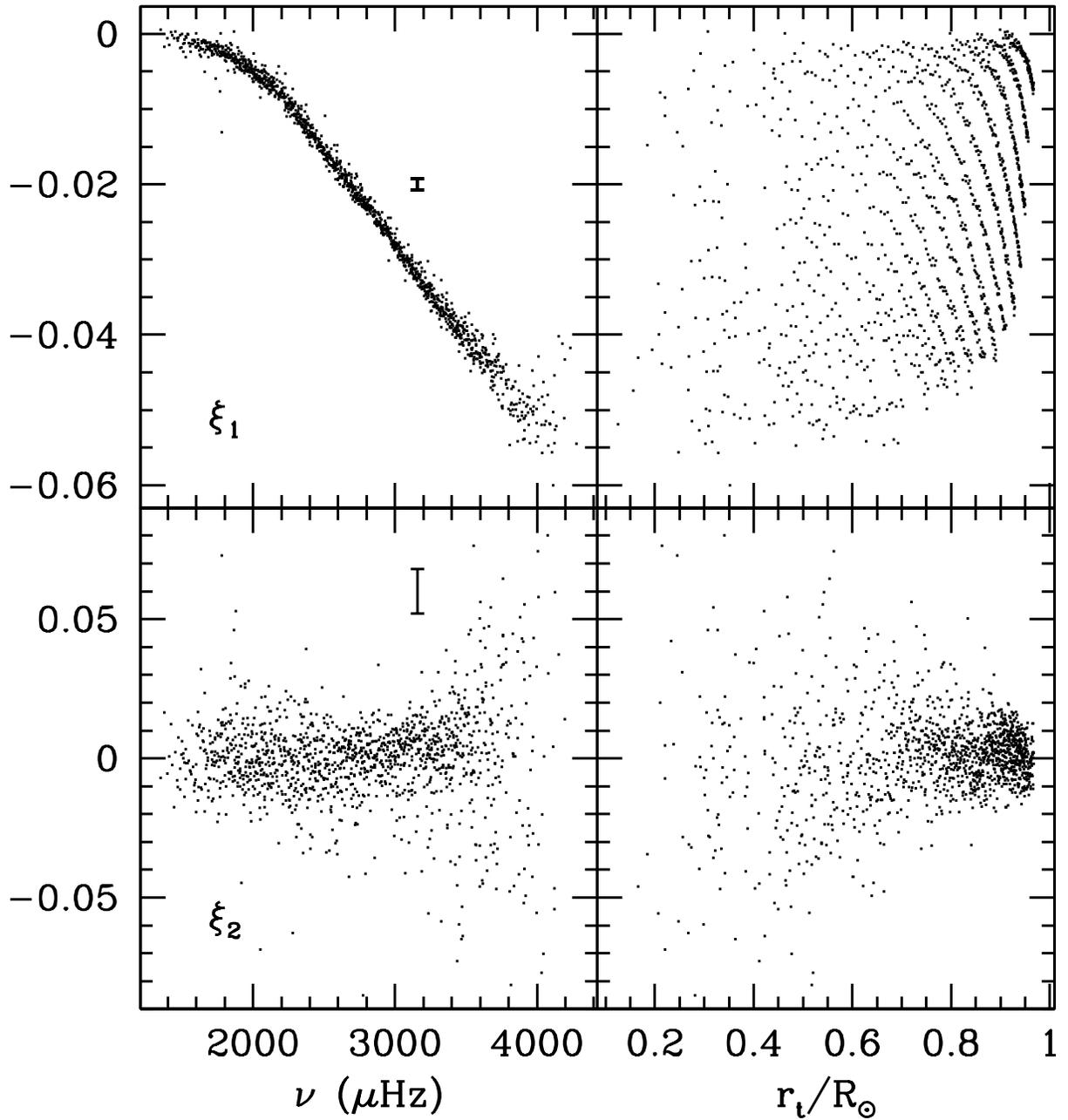}
\caption{The first two eigenvectors for the GONG data sets plotted as 
a function of frequency (the left-hand panels) and as a function of the 
lower turning radius (the right-hand panels).  The first eigenvector $\vec{\xi}_1$ 
shows a signature similar to the one seen in the MDI data, albeit much 
less obviously.  The eigenvector $\vec{\xi}_2$ shows no structure, unlike the 
equivalent eigenvector for MDI, implying that the MDI $\vec{\xi}_2$ is an 
instrumental artifact.
\label{fig:GONGvec}}
\end{figure}

The mode parameters measured from GONG data are somewhat noisier 
than those measured from MDI data.  Nevertheless, a similar 
analysis of the GONG data allows us to confirm results obtained from 
MDI data.  The first two eigenvectors from the GONG observations 
are shown in Fig. \ref{fig:GONGvec}.  As with the MDI data, $\vec{\xi}_1$ plotted 
against $r_t$ shows some structure in the deep interior.  The second 
eigenvector, $\vec{\xi}_2$, shows no more structure than the MDI $\vec{\xi}_3$ 
eigenvector, and the remaining eigenvectors appear to be Gaussian noise, 
reinforcing our conclusions that the structure in $\vec{\xi}_2$ from the MDI data 
set is instrumental in origin.

\begin{figure}
\plotone{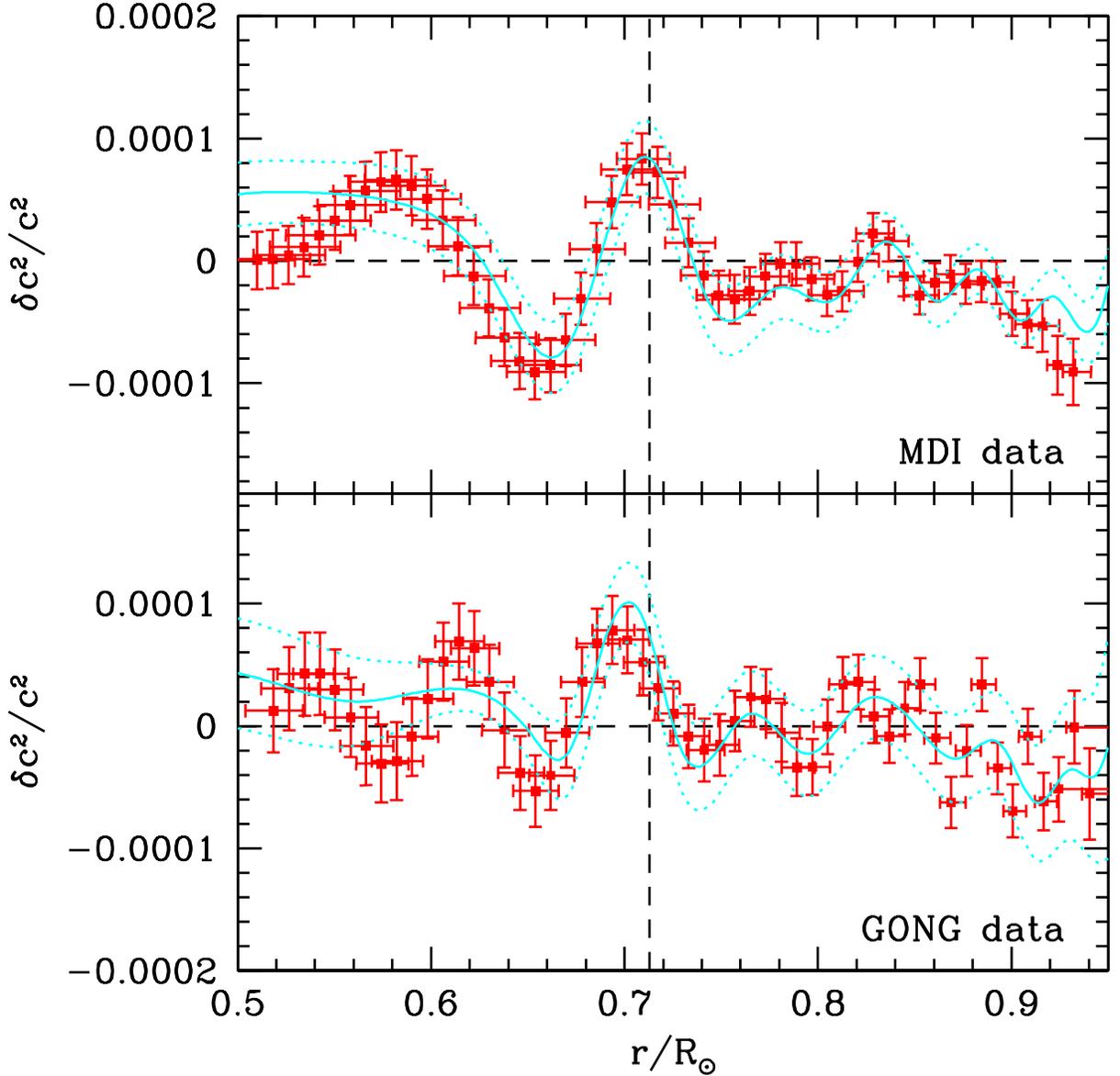}
\caption{Inversion for sound speed of the $\vec{\xi}_1$ eigenvector.  
The top panel shows the inversion of the MDI data.  The lower panel 
shows the inversion of the GONG data.  The solid cyan line is the 
result from the RLS inversion (the dotted lines are the vertical 
error bounds).  The red points are the results from the SOLA 
inversion.  The horizontal dashed line is the zero-point.  The 
vertical dashed line represents the location of the base of the 
convection zone.  At the convection zone base, the MDI inversion 
results show a clear depression in sound speed at high activity 
(the sense of the inversion is low activity minus high activity) 
and an enhancement in the tachocline region.  The depression is 
matched in the GONG inversion results.  The location of this feature, 
though slightly deeper, is within the horizontal errors of the 
MDI result.
\label{fig:MDIinv}
\label{fig:GONGinv}}
\end{figure}

\begin{figure}
\plotone{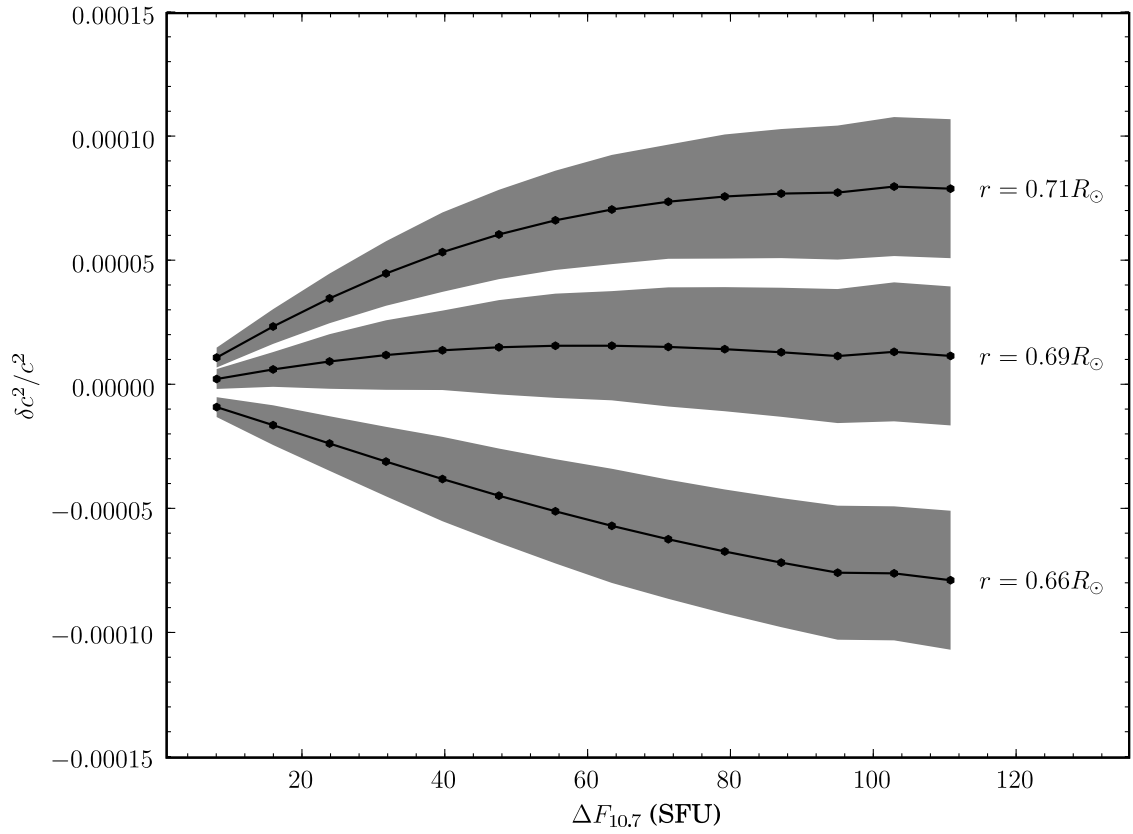}
\caption{Change in inferred sound speed as function of activity level 
(10.7cm radio flux) is shown for different radii around the base of 
the convection zone.  The shaded regions show the errors for each 
set of inversions.
\label{fig:MDIcsq}}
\end{figure}

We invert the appropriately scaled $\vec{\xi}_1$ eigenvectors to determine the 
change in the sound speed as a function of radius.  
We show the results of the inversion of the $\vec{\xi}_1$ vector with the 
\#1216 base set in Fig. \ref{fig:MDIinv}.  This vector has been scaled by 
the coefficient for set \#3160 (start date:  2001 August 27) in order to 
give the inversion results physical meaning.  It 
is readily apparent that at a depth of $r \approx 0.713R_\odot$, near the 
location of the convection zone base, there is a change in the sound speed.  
This feature is well matched in both the RLS and the SOLA inversions, 
which implies that the feature is actually present in the data.  The depression 
in sound speed at the base of the convection zone with increasing activity is 
matched with a corresponding enhancement below the convection zone.  
We invert the GONG data as well.  The inversion results are shown in Fig. 
\ref{fig:GONGinv}.  There is a clear feature at the base of the convection 
zone, as seen with MDI data.  The presence of this feature in 
the data of an independent instrument with an independent reduction pipeline is 
very encouraging --- it strongly implies that the changes implied by the 
inversions are present in the Sun itself rather than artifacts in the data.  
Figure \ref{fig:MDIinv} shows the difference in sound speed between two extrema 
in the solar cycle.  To show how the interior changes with time, in Fig. 
\ref{fig:MDIcsq} we show the sound speed inversions at three radii as a function 
of $\Delta F_\mathrm{10.7}$.

We confirm, therefore, that the signature in mode frequencies is consistent 
with a change in structure at the base of the convection zone.  Other 
authors have looked for changes in this region, but have not found any changes.  
The change that we have detected, while statistically significant, is 
very small, and it is only with the benefit of an entire solar cycle's 
worth of high precision observations that we can detect changes at this level.  
\citet{BA01} examined the mode frequencies for evidence of a 
change in the location of the base of the convection zone.  They did not 
detect any change, and the sound speed profile that we find in Fig. \ref{fig:MDIinv} 
is very different from the one they expected from a change in the base of 
the convection zone.  This implies that, even if the change we are detecting 
is thermal in nature, it is unlikely to be related in any way to a change 
in the position of the base of the convection zone.

These inversions have been done assuming that the frequency differences are a result 
of a change in sound speed only.  It is almost certain, however, given how tightly 
correlated this change is with solar activity, that the observed changes are 
related in some way to changes in the internal magnetic fields.  What we have really 
inverted for, therefore, is a change in the {\em wave} speed.   If we assume 
that the entire change is due to a change in the wave propagation speed due the presence of 
magnetic fields, in other words that $\delta c^2 / c^2 \approx v_A^2 / c^2$, as in 
\citet{Betal04}, we can obtain a value for $B$.  The change at the 
base of the convection zone is $\delta c^2 / c^2 = (7.23 \pm 2.08) \times 10^{-5}$, 
which implies a magnetic field strength of 290 kG.  This is consistent with 
the results of earlier authors --- \citet{GD93} placed an upper limit of 1MG on the 
toroidal field at the base of the convection zone, and \citet{Basu97} found that the 
magnetic field in this region could not exceed 300kG.  \citet{CS02} found somewhat 
stronger fields (400 to 700kG).

\subsection{Latitudinal changes}
\begin{figure}
\plotone{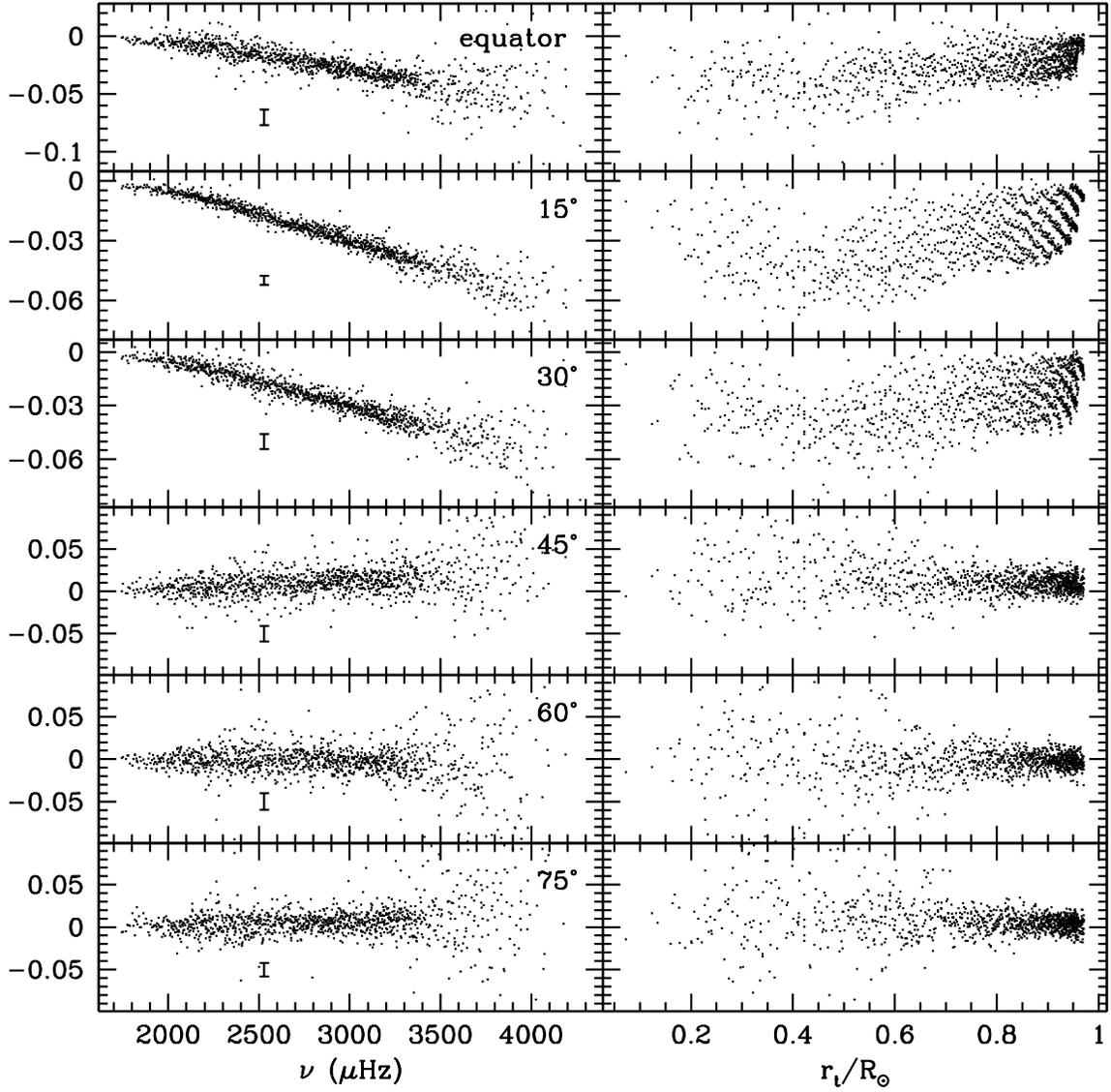}
\caption{The first eigenvector $\vec{\xi}_1(\theta)$ for latitudes from 
0$^\circ$ to 75$^\circ$.  The data are from MDI and are relative to the 
\#1216 mode set.  The left-hand panels show the eigenvectors as functions 
of frequency, and the right-hand panels as functions of the lower turning 
points.
\label{fig:LATvec}}
\end{figure}

\begin{figure}
\plotone{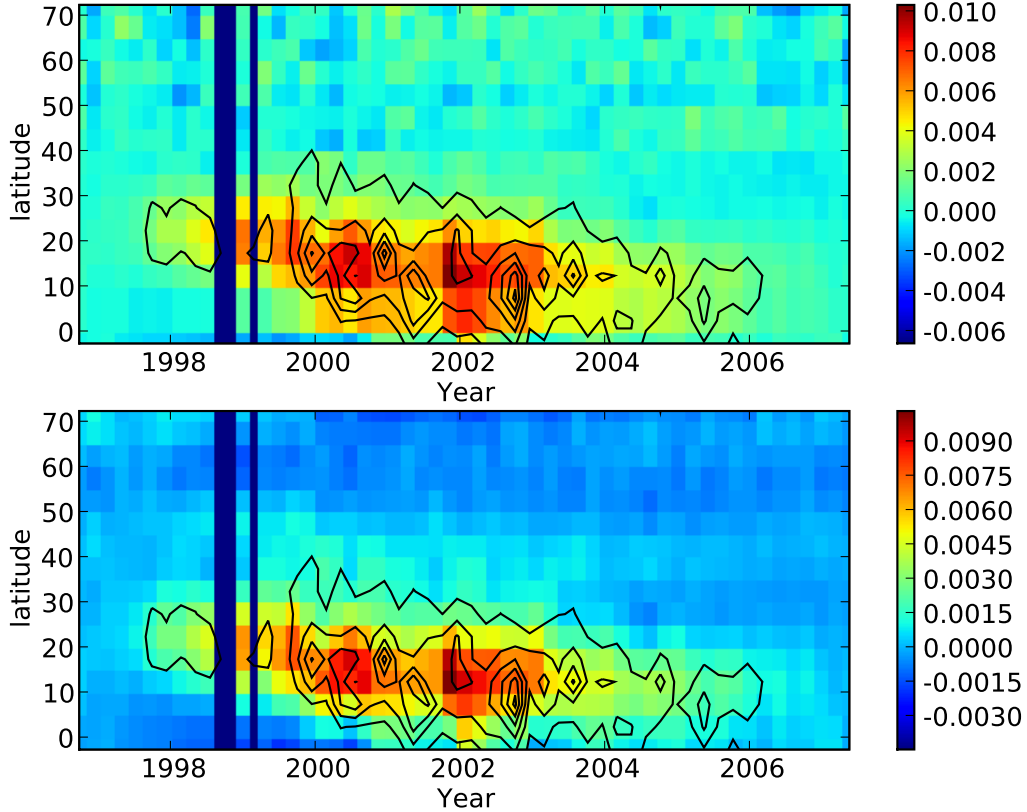}
\caption{The scaling coefficients are plotted as a function of time 
and latitude.  The top panel shows the coefficients for each individual 
eigenvector $\vec{\xi}_1(\theta)$.  The bottom panel shows the scaling coefficients 
for all the latitudes as a function of the $\vec{\xi}_1(15^\circ)$.  This shows 
how the changes represented by that eigenvector change as a function of 
both time and latitude.  The average unsigned magnetic flux from MDI 
carrington rotation synoptic maps over each 72-day period is shown in contour.  
The contours are spaced every 52G, with the lowest at 56.5G.  The vertical 
bars in 1998 are gaps in MDI coverage due to spacecraft problems.
\label{fig:LATcoef}}
\end{figure}

The MDI and GONG data sets also contain splitting coefficients.  The even-order 
coefficients contain information about the non-spherically symmetric structure 
in the solar interior.  Because the surface manifestations of solar activity 
are strongly latitudinally dependent, we have used these coefficients to study the 
temporal variability of structure at different latitudes.  The frequencies 
corresponding to different latitudes are computed using equation \ref{eq:lat}.  
The PCA procedure is performed for each latitude as was done with the mean frequencies, 
and as usual is done with respect to set \#1216.  
The first eigenvector for six different latitudes is shown in Fig. \ref{fig:LATvec}.  
When plotted as a function of frequency, the latitudes from the equator to 30$^\circ$ 
show a similar frequency dependence as in the case of the mean frequencies in 
Fig. \ref{fig:MDIvec}.  When plotted as a function of $r_t$, we see change at 
and below the convection zone base.  The higher latitudes show no 
structure, and the eigenvectors for these latitudes are consistent with Gaussian 
noise.  The scaling coefficients for each latitude as a function of time are 
shown in Fig. \ref{fig:LATcoef}, along with the surface magnetic field.  Like 
the scaling coefficients for the mean frequencies shown in Fig. \ref{fig:MDIcoef}, 
the latitudinal scaling coefficients closely follow the surface activity.

\begin{figure}
\plotone{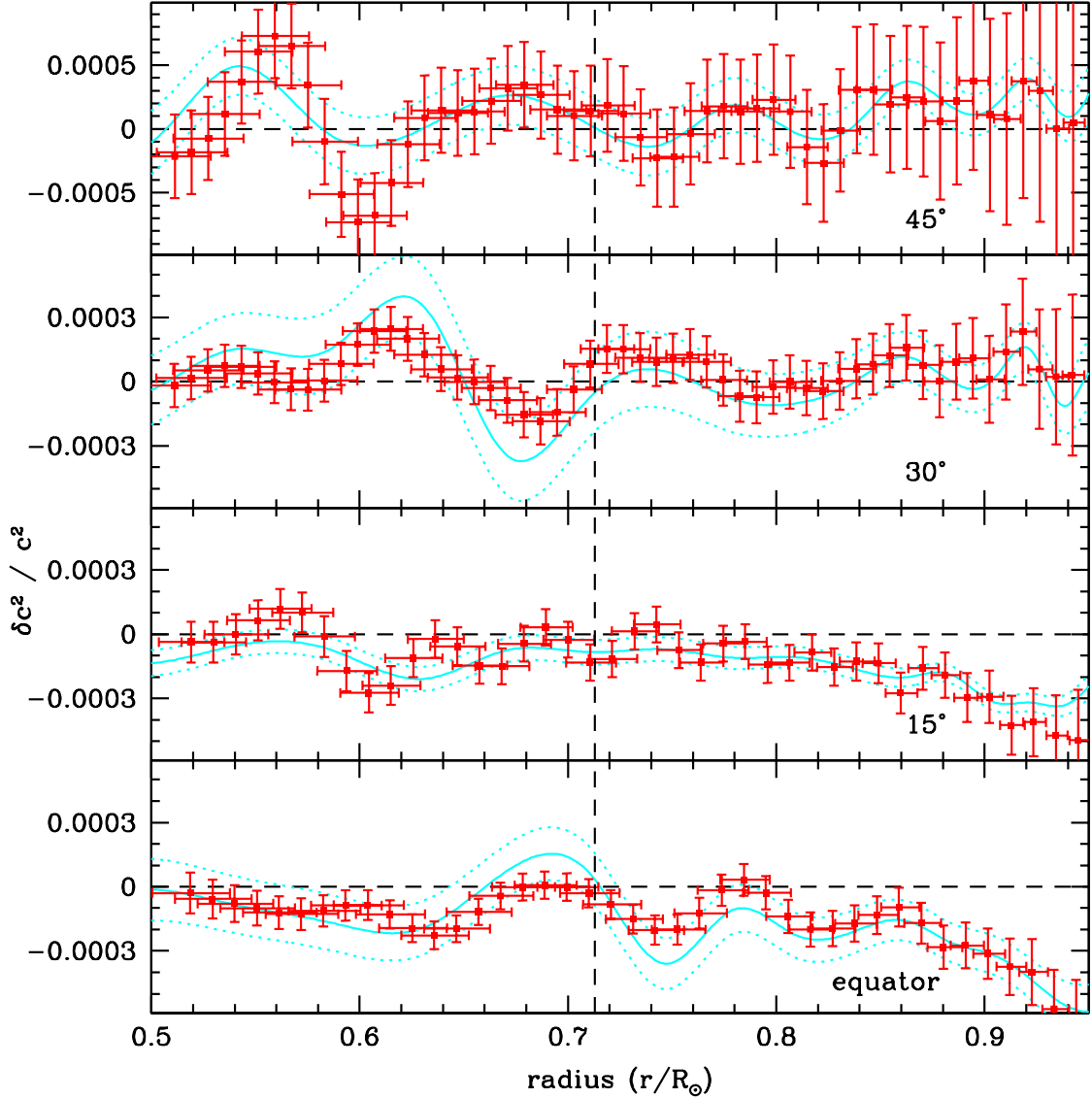}
\caption{The inversions of the latitudinal frequencies are shown for 
four different latitudes from the equator to $45^\circ$.  The differences 
are with respect to MDI set \#1216.  As expected from an examination 
of the raw frequencies, there are no discernible features in the inversion 
for $45^\circ$ and above --- the inversions are extremely unstable to the 
choice of inversion parameters.  At $15^\circ$ and the equator, a significant 
(about 2$\sigma$) enhancement in sound speed at high activity is observed 
above approximately $r=0.98R_\odot$.  As before, the vertical dashed line 
indicates the position of the base of the convection zone ($r=0.713R_\odot$).
\label{fig:LATinv}}
\end{figure}

We show the sound speed inversions for the equator, 15$^\circ$, 30$^\circ$, 
and 45$^\circ$ in Fig. \ref{fig:LATinv}.  The errors in the eigenvectors are 
larger here than for the mean frequencies, in large part because each frequency 
is a combination of mean frequency and splitting coefficients, each with their own 
errors.  Nevertheless, there are several points of interest 
in these inversions.  The first is a clear sound speed change for radii greater than  
approximately $r=0.86R_\odot$ at 15$^\circ$ and the equator.  The change seen 
in the SOLA inversion results is well matched in this region by the RLS inversion results, 
and the significance of the change approaches 2$\sigma$.  There is the 
possibility, though less statistically significant, of a change at greater depth, 
i.e., approximately $r=0.82R_\odot$.  At 30$^\circ$, a change in sound speed through 
the tachocline is seen in the RLS results, but it does not appear to be as clear 
in the SOLA results..  It is unclear whether or not this result is statistically 
significant.  At latitudes of 45$^\circ$ and higher, the inversion errors become large, 
and the inversion results themselves become extremely sensitive to the inversion parameters.  
We conclude that there are no structural changes in the solar interior at 
or above a latitude of 45$^\circ$ large enough to be present in our data sets.

\begin{figure}
\plotone{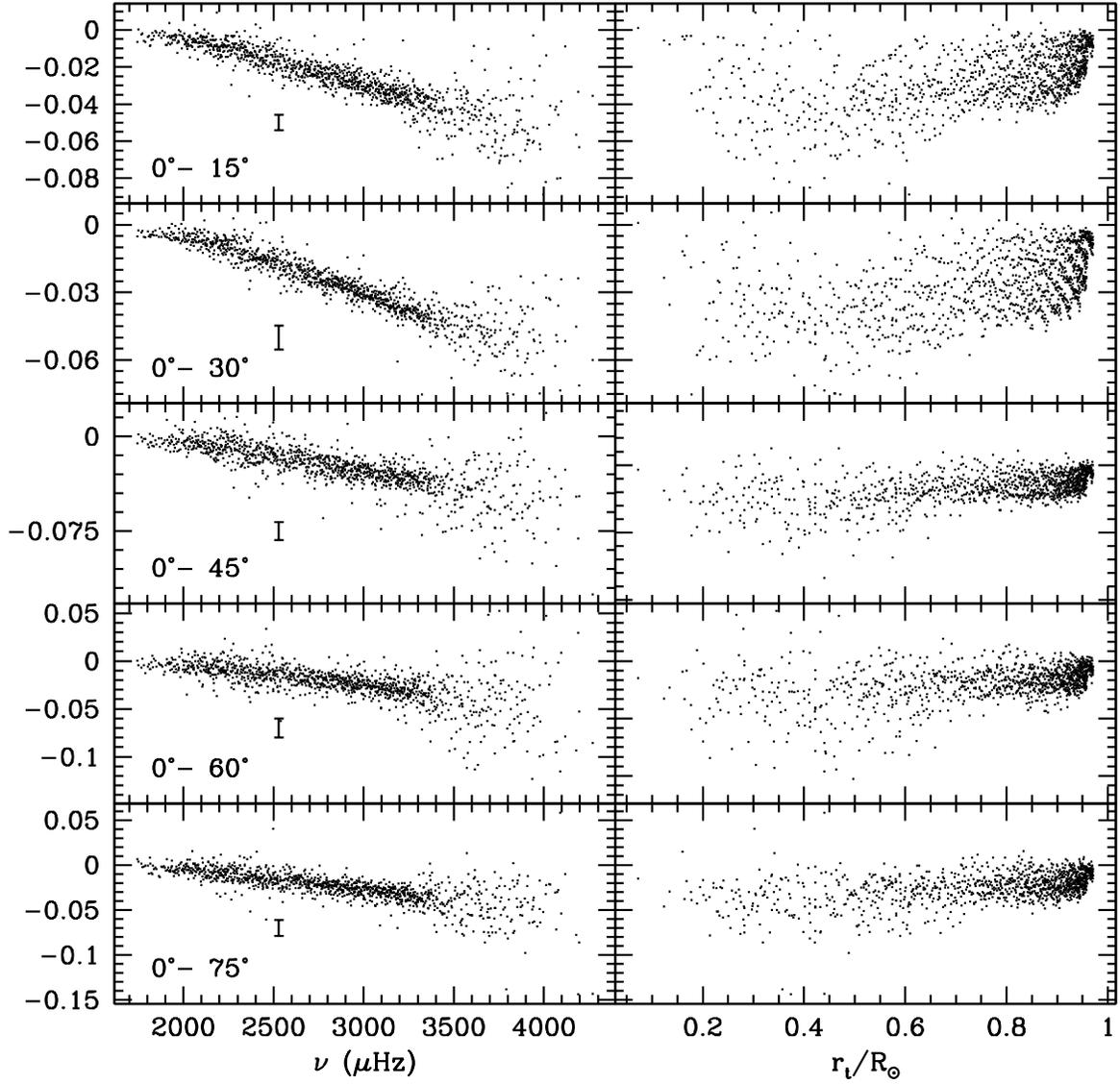}
\caption{The $\vec{\xi}_1$ eigenvectors for the asphericity terms --- equator minus 
latitude --- are shown for five different latitudes.  As usual, the left panels 
show the eigenvectors with respect to frequency, and the right hand panels are 
with respect to lower turning point.
\label{fig:ASPHvec}}
\end{figure}

\begin{figure}
\plotone{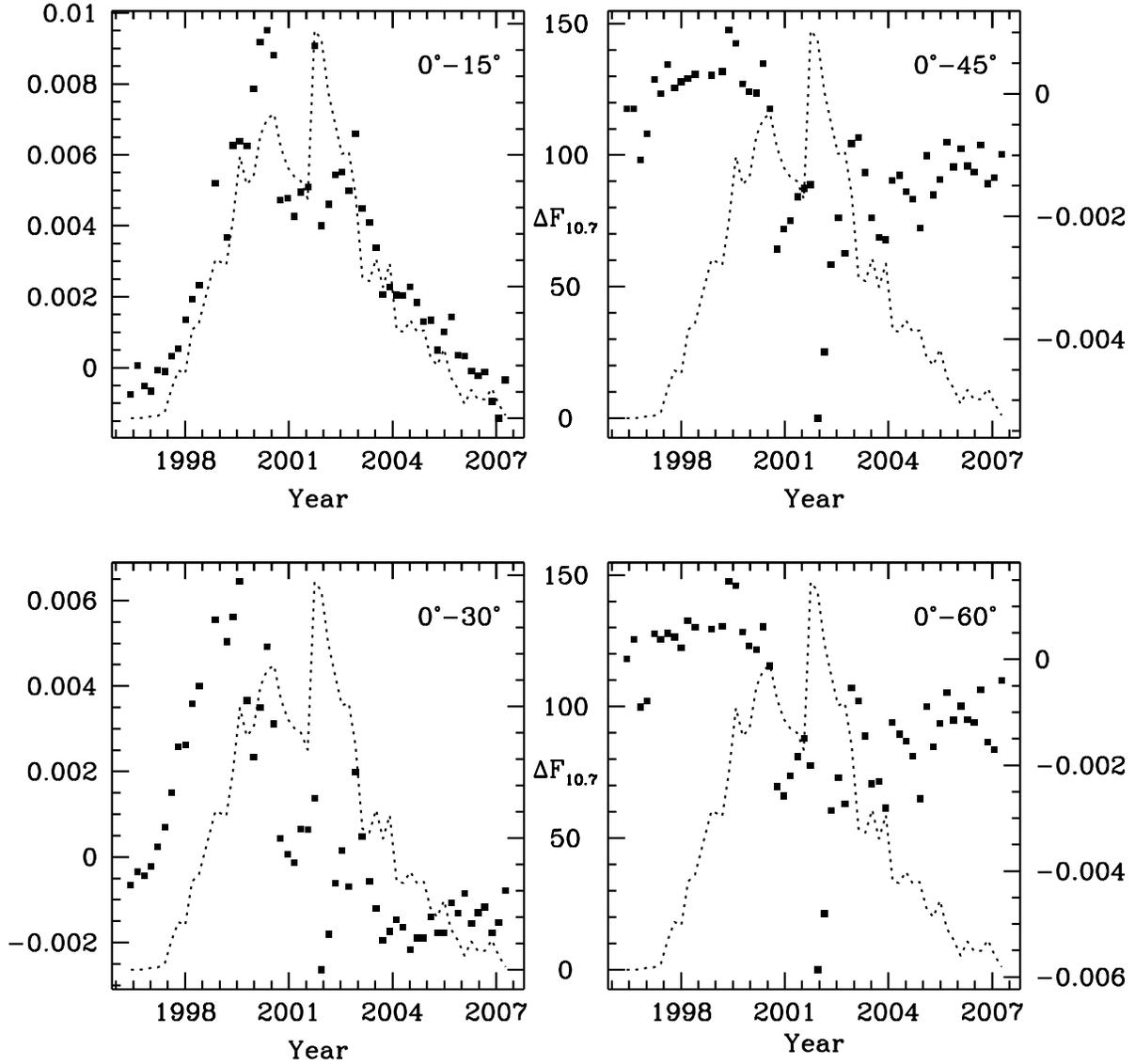}
\caption{The scaling coefficients for the asphericity eigenvectors are shown.  
The dotted line is the $F_{10.7}$ flux.  The $0^\circ-15^\circ$ and $0^\circ - 30^\circ$ 
both show a phase shift, consistent with the butterfly diagram from Fig. \ref{fig:LATcoef}.  
The higher latitudes show a distinct change from mostly positive to negative at 
high activity.
\label{fig:ASPHcoef}}
\end{figure}

We have also analyzed directly the change in solar structural asphericity over 
the course of the solar cycle, by taking the scaled frequency differences of 
high latitudes with respect to the equator.  For this analysis, therefore, we have 
54 mode sets at each latitude.  The difference is equator$-$latitude.  
Figure \ref{fig:ASPHvec} shows the 
$\vec{\xi}_1$ eigenvectors for five different latitudes with respect to the equator.  
Clearly, the signal to noise in these data are worse than either the latitudinal 
frequencies or the mean frequencies, but radial structure is discernible in the 
eigenvectors for the equator relative to 15$^\circ$, 30$^\circ$, and 45$^\circ$.  
In fig. \ref{fig:ASPHcoef}, the scaling coefficients for the asphericity terms 
are shown.  There is an evident phase delay in the scaling coefficients of 15$\circ$ 
and 30$^\circ$ with respect to the equator.  This is expected from Fig. \ref{fig:LATcoef}.  
The scaling coefficients for higher latitudes (45$^\circ$ and 60$^\circ$ with respect 
to the equator) show no structure except for an abrupt change from mostly positive 
to negative, corresponding closely to the peak of the solar activity cycle.

\section{CONCLUSIONS}
\label{sec:conc}
We have analyzed the changes in solar oscillation frequencies over the 
course of solar cycle 23.  In order to reduce the effects of 
measurement errors and detect the faintest signatures of solar 
variability possible, we have employed a Principal Component Analysis 
of the frequencies.  The mean frequencies are known to vary over the 
solar cycle, and this variation is known to be tightly correlated with 
surface activity.  We have found this correlation as well.  

In addition to the frequency dependent change which these 
earlier authors have detected, we have found a small but statistically 
significant change in modes with turning points at or below the convection 
zone.  This confirms the result of \citet{CS05}.  This 
signature is present in both the MDI and the GONG data sets.  We have 
inverted these results to obtain the difference in sound speed, and we 
have confirmed that there is a change in solar structure at the base of 
the convection zone over the course of the solar cycle.  The measured 
change at a radius of $r=(0.712^{+0.0097}_{-0.0029})R_\odot$ is 
$\delta c^2 / c^2 = (7.23 \pm 2.08) \times 10^{-5}$, where 
the errors in radius are a measure of the resolution of the inversion 
taken from the first and third quartile points of the inversion kernel.

We have also used the splitting coefficients to investigate how the 
changes in structure vary over latitude.  We have found that the 
changes in the solar interior are tightly correlated with the 
latitudinal distribution of surface activity.  The most statistically 
significant changes detected in the analysis are changes in sound 
speed in the outer ten percent (by radius) of the solar interior.

\acknowledgements
We wish to thank the anonymous referee for helpful comments which helped 
to clarify a number of the discussions in this paper.  We are also grateful 
to T.~ Larson and J.~ Schou for their helpful discussion of the anomalous 
results in the MDI data.  
This work  utilizes data obtained by the Global Oscillation
Network Group (GONG) project, managed by the National Solar Observatory,
which is operated by AURA, Inc. under a cooperative agreement with the
National Science Foundation. The data were acquired by instruments
operated by the Big Bear Solar Observatory, High Altitude Observatory,
Learmonth Solar Observatory, Udaipur Solar Observatory, Instituto de
Astrofisico de Canarias, and Cerro Tololo Inter-American Observatory.
This work also utilizes data from the Solar Oscillations
Investigation/ Michelson Doppler Imager (SOI/MDI) on the Solar
and Heliospheric Observatory (SOHO).  SOHO is a project of
international cooperation between ESA and NASA.
MDI is supported by NASA grants NAG5-8878 and NAG5-10483
to Stanford University. This work is partially supported by NSF grants 
ATM 0348837 and ATM 073770 to SB.

\appendix
\section{PRINCIPAL COMPONENT ANALYSIS}
\label{sec:PCAappend}
We give a brief description of the Principal Component Analysis 
technique that follows the one found in \citet{Kendall80}.  This technique assumes 
that we begin with a set of observations, each one consisting of a set of data 
points (for example, a set of spectra of different objects, or a set of images, or, 
as in this paper, sets of mode frequency measurements at different points in time).  
Assume we have 
$m$ observations, each with $n$ data points.  The $m$ vectors $\vec{x}_i$ each 
contain the $n$ data points $x_{ij}$, measured relative to the mean $\bar{\vec{x}}_j$
We wish to find a new set of vectors $\vec{c}_i$ that are linearly dependent 
on $\vec{x}_i$, and uncorrelated with each other.  In addition, we require that they 
have stationary values of their variance.  This condition is imposed to ensure that 
most of the variance is accounted for in as few vectors $\vec{c}_i$ as possible.  
Alternatively, this condition can be viewed geometrically as a rotation of the basis 
vectors such that the variance of the data with respect to each basis vector is maximal.  
This is most easily understood in the case of a two dimensional data set, where PCA is 
equivalent to a linear fit to the data, and the rotation sets one basis vector parallel 
to that fit.  
The vectors $\vec{c}_i$ will be given by a linear combination of the original observations:
\begin{equation}\label{eq:AP1}
\vec{c}_i = \sum_{j = 1}^n \xi_{ji}\vec{x}_j.
\end{equation}
The vectors $\vec{\xi}_j$ will form a new basis set for the observations.  The variance of 
$\vec{c}_i$ is given by
\begin{equation}\label{eq:var}
\textrm{var}\ \vec{c}_i = \sum_{j=1}^n \sum_{k=1}^n \xi_{ji} \xi_{ki} e_{jk},
\end{equation}
where $e_{jk}$ is the covariance between $x_j$ and $x_k$.  The covariance matrix is given
by
\begin{equation}\label{cov}
\mathbf{E} = \frac{1}{n}\mathbf{XX}^\mathrm{T},
\end{equation}
where $\mathbf{X}$ is the matrix $\left( \vec{x}_1, \vec{x}_2, \ldots, \vec{x}_m \right)$.  In order to 
get unique solutions for the values of $\xi_{ji}$, we impose a normalization condition:
\begin{equation}\label{eq:constraint}
\sum_{i=1}^m \xi_{ji}^2 = 1.
\end{equation}
Finding the stationary values of $\textrm{var}\ \vec{c}_i$ (eq. \ref{eq:var}) with the 
condition \ref{eq:constraint} is equivalent to finding the stationary values of:
\begin{equation}\label{eq:newvar}
\sum_{j=1}^m \sum_{k=1}^m \xi_{ij} \xi_{ik} e_{jk} - \lambda \left( \sum_{j=1}^m \xi_{ij}^2 - 1 \right),
\end{equation}
where $\lambda$ is some constant.  To find 
the stationary values, we differentiate \ref{eq:newvar} by $\xi_{ij}$ and find the roots 
of the equation:
\begin{equation}
\sum_{k=1}^m \xi_{ik} e_{jk} - \lambda \xi_{ij} = 0.
\end{equation}
This is an eigenvalue problem --- the vector $\vec{\xi}_j$ is an eigenvector of
$\mathbf{E}$ and $\lambda$ is the corresponding eigenvalue.  The eigenvectors form an orthonormal 
basis set and are called the `principal components' of the data set.  When they are
ordered by decreasing eigenvalue $\lambda$, the first principal component will have 
the largest possible variance, the second the second largest variance, and so on.  In 
our work, we use the singular value decomposition of the covariance matrix to get the 
eigenvalues and vectors.  The vectors $\vec{c}_i$ are the scaling coefficients for the 
principal components.

The PCA technique has several known weaknesses.  The first is that its simplest
implementation requires a completely filled covariance matrix --- in other words
there can be no missing data in $\mathbf{X}$.  This is a problem for us  
since not all modes are identified
in each different observation epoch.  There are generalizations
of the technique which allow for large quantities of missing data.  In our case, however,
the modes in which we are interested tend to be identified in most of the mode sets, so
we can simply interpolate the missing modes from the existing frequency measurements 
without any significant effect on the results.  We do not, therefore, use a more 
specialized technique.

A second problem with PCA is its sensitivity to outliers.  One often-quoted example 
shows a PCA decomposition where the correlation coefficient between the first two 
components with one outlier removed is 0.99 \citep[e.g.,][]{Huber81}.  There are 
available routines for making PCA more robust, but in our case, because the 
dimensionality of the problem is relatively small, we can instead empirically test 
the sensitivity of the data set to outliers.  Monte Carlo tests show that our data 
set is not prone to errors in the PCA due to outliers.

Principal Component Analysis has been used in a wide variety of astronomical contexts 
(see references in \citealp[Chapter2]{MH87}).  In solar physics, PCA has been used for, 
among other things, the inversion of Stokes profiles \citep[e.g.,][]{Eydenbergetal05,R-Vetal06}, in 
detecting structures in coronal activity \citep[e.g.,][]{Cadavidetal07}, and in helioseismic 
rotation inversions \citep[e.g.,][]{E-Detal04}.

\end{document}